\documentstyle[aps,prl]{revtex}
\begin{document}

\twocolumn[\hsize\textwidth\columnwidth\hsize\csname
@twocolumnfalse\endcsname

\draft
\author{Gonzalo Usaj$^1$, Horacio M. Pastawski\thanks{Corresponding author, Email: horacio@famaf.fis.uncor.edu}$^{1,2}$, and
Patricia R. Levstein$^1$}
\address{$^{1}$Facultad de Matem\'{a}tica, Astronom\'{\i }a y F\'{\i }sica, Universidad 
Nacional de C\'{o}rdoba, Ciudad Universitaria, 5000 C\'{o}rdoba, Argentina}
\address{$^{2}$International Center for Theoretical Physics, P. O. Box 586, 34100 
Trieste, Italy}
\title{Dynamically Driven Irreversibility in Many Body Quantum Systems: An
experimental approach.}
\date{Received December 29, 1997}
\maketitle

\begin{abstract}
The NMR technique allows one to create a non-equilibrium local polarization
and to detect its later evolution. By a change of the sign of the effective
dipolar Hamiltonian, the {\em apparently} diffusive dynamics is reverted,
generating a polarization echo (PE). This echo attenuates as a function of
the time $t_R$ elapsed until the dynamics is reverted. We report a set of
experiments where we slow down the dipolar dynamics showing that it controls
this irreversibility. In cobaltocene, a cross over from Gaussian to
exponential PE\ attenuation is found; in ferrocene the exponential regime is
not reached. We conclude that many-body quantum systems have intrinsic
instabilities, which amplify the environment fluctuations and hinder their
backward evolution.
\end{abstract}

] \narrowtext

This letter reports the results of a set of NMR experiments designed to show
the relationship between the complex nature of many-body dynamics and
irreversibility. Since Boltzmann's introduction\cite{Boltzmann} of the
``Stosszahlansatz'' the understanding of why an initial ordered state {\em \ 
}{\it evolves irreversibly} towards a higher entropy state has been a
challenging issue\cite{lebo}. Loschmidt observed\cite{Loschmidt} that since
microscopic laws are reversible, it should be possible to get a backward
evolution recovering the initial state. However, this has remained in the
realm of ``gedanken'' experiments, task of an imaginary Loschmidt's daemon,
for almost a century. Recent theoretical approaches to understand
irreversibility point to the concept of {\it local instabilities} associated
with classical chaotic systems\cite{Izrailev}. If some initial state evolves
according to a microscopic law for a time $t_{R}$ and then one makes the
system evolve backwards, at time 2$t_{R}$ the state differs from the
original one by a ``distance'' proportional to $\exp \left[ \gamma
t_{R}\right] .$ The Lyapunov exponent $\gamma $ characterizes the chaotic
motion responsible for the amplification\cite{Bellemans} of any small error
in the evolution. However, the quantum one-body counterpart\cite{Izrailev},
does not seem to present this instability leaving a conceptual gap in the
formulation of a quantum dynamical description of irreversibility.
Analytical or numerical studies of many body quantum systems are lacking.

Experimentally, a breakthrough into irreversibility was set up by E. Hahn
with his NMR\ Spin Echo\cite{Hahn} which also applies to other spectroscopic
techniques. It uses that the evolution of the macroscopic polarization of a
system of $N$ noninteracting spins, $\{I_i\},$ is described by the
superposition of the {\it one body} Hamiltonians ${\cal H=}\sum_i{\cal H}_i.$
An external control parameter allows one to switch ${\cal H}_i\rightarrow -%
{\cal H}_i$ leading to a reversion of the dynamics. In this case, eventual
many body interactions limit the reversibility. Many body dynamics seemed
much more difficult to invert until Rhim et al. \cite{Rhim} achieved a
Loschmidt's daemon for dipolar coupled spins. Since the sign and magnitude
of this interaction depends on the orientation of spins relative to their
internuclear vector, the switch ${\cal H}\rightarrow -{\cal H}$ is possible
by changing the quantization axis. A further development is due to Zhang,
Meier and Ernst\cite{ZME}\ who were able to create and detect the local
polarization of a spin $I_{1,}$ represented by the density matrix ${\bf \rho 
}_0=(\frac 12{\bf 1}+I_1)/2^{N-1}$. This allows one to deal with a well
defined microstate\cite{PLU,PUL}. The recovered polarization after a forward
and backward evolution is called Polarization Echo (PE). Later experiments%
\cite{LUP} suggested that there are systems where the PE amplitude $M_{PE}$
decays exponentially with time $t_R$, while in others it decays with the
much stronger Gaussian law. Figure 1a shows an idealized schematic view of
the PE experiment. The initial state ${\bf \rho }_0$ evolves with ${\cal H}%
+\Sigma ,$ where $\Sigma $ describes small interactions not contained in the
truncated dipolar Hamiltonian ${\cal H}$. The dashed curve in Fig. 1 is the 
{\em local polarization} detected at time $t.$ It decays as $1-\overline{d}%
^2t^2/2\hbar ^2+{\cal O}(t^4)$, with $\overline{d}^2$ the dipolar second
moment, evidencing an evolution according to the quantum many body dynamics%
\cite{QZeno}. This polarization spreads beyond the neighborhood of site $1$
in a typical time $\tau _{mb}$ which is a few times $\hbar /\overline{d}$.
If at a time $t_R$ we change the invertible part of the Hamiltonian ${\cal H}%
\rightarrow -{\cal H}$, a PE builds up with a maximum ($M_{PE}$) at $2t_R$.
The presence of uncontrolled processes prevents the perfect recovery of the
initial state ($M_{PE}<1$). Those ``irreversible'' interactions are simply
the ones we do not know how to control. It is clear that if important
interactions with a thermal bath were present they would show up as an
exponential decay (Fig. 1) and $\Sigma =\Delta -{\rm i}\Gamma $. In this
case the time $\tau _\phi \approx \hbar /|\Gamma |$ is much shorter than $%
\tau _{mb},$ and the irreversibility is directly controlled by the
environment fluctuations. {\em Our hypothesis} is that, when $\tau _{mb}\ll
\tau _\phi $, the complex structure of the many body dynamics ($N\gg 1$)
plays a crucial role in the attenuation of the PE. Then, intrinsic
instabilities of the many body dynamics would lead to a {\em stronger}
Gaussian decay (Fig. 1). This could explain the amplification of
fluctuations that, otherwise, could not produce an effective relaxation on
the observed time scale.

In order to test this hypothesis one must {\em control the complexity} of
the state reached at time $t_R$. This can be done in two ways: by changing
the system (and hence the interaction network and relaxation mechanisms) or
by reducing the effective dynamics. The latter provides a more controllable
situation. Thus, in addition to selecting appropriate systems, we developed
a pulse sequence which allows us to progressively reduce the complexity
reached by the many body state, while the non-invertible interactions are
kept constant. This is achieved by fitting $n$ periods of alternating {\em %
forward} and {\em backward} dipolar evolutions in a fixed time{\em \ }$t_R$.

The sequence is schematized in Fig.2. It uses the rare $^{13}$C spin ($S$%
-spin) (1.1 \% natural abundance) as a {\em local} probe that injects
magnetization to one of the abundant $^1$H spins ($I$-spins) and later
captures what is left. A $\left( \pi /2\right) _x$ pulse on the $I$-spins
creates a polarization that is transferred during $t_C$ to the $S$-spin when
both are irradiated at their respective resonant frequencies with field
strengths fulfilling the Hartmann-Hahn condition. After the $S$-spin is
polarized, it is kept spin-locked for a time $t_S$ while the coherence of
the $I$-spins decays to zero. Then: {\bf A) }A cross polarization (CP) pulse
of duration $t_d$ transfers the magnetization from an initially $\widehat{y}$
polarized $S$-spin to the $y$-axis of the $I_1$-spin directly bonded to it%
\cite{Muller}, establishing the initial state ${\bf \rho }_0$; {\bf B) }The
dipolar coupled $I$-spins evolve in presence of a strong spin-lock field
that fixes the quantization axis and prevents the $I$-$S$ coupling. If we
neglect non-secular terms, the effective Hamiltonian 
\begin{equation}
{\cal H}^y=\left[ -\frac 12\right] \sum_{j>k}\sum_kd_{jk}\,\left[
2I_j^yI_k^y-\frac 12\left( I_j^{+}I_k^{-}+I_j^{-}I_k^{+}\,\right) \right] ,
\label{Hy}
\end{equation}
governs the {\em forward} evolution of ${\bf \rho }_0$. The flip-flop terms $%
I_j^{+(-)}I_k^{-(+)}$ cause the spreading (``diffusion'') of the initial
localized state; {\bf C}){\bf \ }A $\left( \pi /2\right) _x$ pulse tilts the
polarization to the laboratory frame where the $I$-spins evolve with $%
-\left[ 2\right] {\cal H}_{}^z$ while an $S$-irradiation prevents again the $%
I$-$S$ coupling. $\,$A $\left( \pi /2\right) _{-x}$ pulse tilts the
polarization back to the rotating $xy$ plane. Taking the $\left( \pi
/2\right) $ pulses into account, the effective Hamiltonian in this period is 
$-\left[ 2\right] {\cal H}_{}^y${\bf . }This change in the sign produces the 
{\em backward} evolution that builds up the PE. For a given time $t_R$, we
alternate $n$ periods of forward and backward evolutions with durations of $%
\tau _1=t_R/n$ and $\tau _2=\left[ \frac 12\right] \tau _1$ respectively. 
{\bf D)} Another CP pulse of length $t_d$ is applied to transfer back the
polarization to the $x$-axis of $S$. {\bf E) }The $S$ polarization is
detected while the $I$-system is kept irradiated (high-resolution
condition). The $S$-signal intensity is proportional to the remaining
polarization in the $I_1$-spin after the evolution periods. Since the $S$%
-spin is not an ideal local probe\cite{PLU,PUL}, the proton spins have some
forward evolution during the CP periods. To compensate for this undesired
effect, it is necessary to introduce extra backward periods $\tau ^{\prime
}=\left[ \frac 12\right] t_d/2$. In an ideal situation, the initial state
should be recovered at the end of each cycle. The spreading of the local
excitation due to the dipolar interaction and the number of dipolar spin
correlations can be gradually reduced by increasing $n$. This sequence could
be adapted to a broad range of polycrystals by complementing the magic angle
spinning of the sample with rotor-synchronized pulses\cite{Tomaselli}.

All the NMR measurements were made at room temperature in a Bruker MSL-300
spectrometer, equipped with a standard Bruker CP-MAS probe, operating at a $%
^{13}$C frequency of approximately 75.47 ${\rm MHz}$. We analyzed two
different systems: a single crystal sample of ferrocene\cite{Seiler}, (C$_5$H%
$_5$)$_2$Fe, and a polycrystalline sample of cobaltocene\cite{Antipin}, (C$%
_5 $H$_5$)$_2$Co. The (C$_5$H$_5$)$_2$Fe experimental data were recorded for
one of the two crystallographically inequivalent sites, where the fivefold
molecular symmetry axis is at 20$^{\circ }$ with respect to the external
magnetic field. The (C$_5$H$_5$)$_2$Co data correspond to molecules with the
symmetry axis perpendicular to the external field (they are frequency
selected\cite{LUP} from the $^{13}$C spectrum). The relative phase and the
rf amplitude of the four basic channels ($X$,$-X$,$Y$,$-Y$) are quite
critical and must be carefully adjusted\cite{Burum}.

Fig. 3 shows $M_{PE}$ for the ferrocene sample as a function of $t_R$ for $%
n=1,2,8,16$. We observe that its attenuation decreases with $n$ and, in all
the cases, the {\em dynamical irreversibility} manifests as a Gaussian
decay. We obtain an excellent fitting (solid lines) with the characteristic
time $\tau _{mb}^n$ as the only free parameter (according to \cite{LUP}, we
required $M_{PE}(t_R=0)=1$ and an asymptotic value of $0$). The inset shows
that the dynamical time $\tau _{mb}^{}$ has a linear dependence on $n$. The
good signal to noise ratio in the single crystal for $n=1$ allows us to
ensure the Gaussian decay for over two orders of magnitude. This is a
striking result since, from conventional wisdom, one could expect the
attenuation of the PE to be directly caused by the non inverted
interactions. These residual interactions, whatever they are, act for the
same time $t_R$ independently of the value of $n$. Consequently, it is the 
{\em reversible} dipolar interaction the one which controls the decay of the
PE in ferrocene.

The situation of the cobaltocene sample is rather different. Fig. 4 shows $%
M_{PE}(t_R)$ for $n=1,2,5,8,12$. A crossover is evident between a dominant
Gaussian attenuation to an exponential one. This can be explained as
follows: While cobaltocene has the same interaction network as ferrocene,
and therefore a similar dipolar dynamics, it possesses a strong local source
of relaxation (the paramagnetic Co atom) with a characteristic time $\tau
_\phi $. When $n=1,2$ the evolved state is complex enough, $\tau
_{mb}^{}<\tau _\phi $, and the dominant effect is the dynamical
amplification. However, when the dynamics is sufficiently reduced ($%
n=8,12,16 $) one gets $\tau _{mb}^n>\tau _\phi $ and the irreversible
mechanism becomes dominant. An asymptotic pure exponential decay with $\tau
_\phi \simeq \left( 970\pm 20\right) \mu s$ is reached upon further increase
of $n$ beyond 12$.$ This paradigmatic example shows how the dynamical
instabilities can be slowed down until the underlying mechanisms beyond our
control are revealed. The crossover is not observed in ferrocene, since the
time scale of the corresponding exponential is beyond the experimental
availability. The solid lines in Fig. 4 represent fittings to the equation $%
M_{PE}(t_R)=\exp \left[ -\frac 32t_R/\tau _\phi -\frac 12\left( t_R/\tau
_{mb}^n\right) ^2\right] $. Based on the results obtained for ferrocene we
set $\tau _{mb}^n=an+b$. In this way the whole set of data was fitted with
only three free parameters.

These experimental results may have important consequences in further
developments of a theory of chaos in quantum many body systems. To see this,
let ${\bf \rho }_{{\cal H}+\Sigma }$ be the density matrix at time $t_R$
obtained from the evolution of ${\bf \rho }_0$ with a generalized master
equation\cite{MasterEq} or the Keldysh equation\cite{Keldysh} for the field
operators. Consider that further (backward) evolution is unitary, $U_{{\cal H%
}}=\exp (-i{\cal H}t_R/\hbar ).$ The normalized PE amplitude can be readily
obtained:

\begin{eqnarray}
M_{PE}(2t_R) &=&2{\rm Tr}\left( I_1U_{-{\cal H}}{\bf \rho }_{{\cal H+}\Sigma
}U_{-{\cal H}}^{\dagger }\right)  \nonumber \\
&=&2^N{\rm Tr}\left( {\bf \rho }_{{\cal H}}{\bf \rho }_{{\cal H+}\Sigma
}\right) -1.
\end{eqnarray}
The last expression can be thought as the overlap between two states evolved
from the same initial state ${\bf \rho }_0$ with different Hamiltonians, $%
{\cal H+}\Sigma $ and ${\cal H}$. From this point of view, the logarithm of $%
M_{PE}$ has a suggestive resemblance to some entropy definitions\cite
{entropy}. In semiclassical cases, an exponential decay is expected. In the
non trivial case its rate $\gamma \gg 1/\tau _\phi $ depends only on ${\cal H%
}$, and it is a manifestation of a strong sensitivity to slight changes in
the dynamical parameters. This property is itself a signature of the
presence of chaos\cite{Haake}. However, our experimental result in a quantum
many body system shows that the decay follows the much stronger Gaussian
decay. This could be a consequence of the {\em progressive availability} of
a huge Hilbert space made possible by the dynamics.

The main conclusion of our experiments is that the complex dynamical
structure presented by the many-body quantum systems has a very strong
intrinsic instability towards irreversibility. Probably this instability was
foreseen by the Boltzmann's deep physical insight. While we have a knob to
slow down the many body dynamics, whatever is left warrants an irreversible
decay. Once we reach the time scale of a process we cannot control it
appears as an ``ultimate'' irreversible time.

HMP acknowledges a fruitful discussion with Prof. E. L. Hahn at a Gordon
Conference. HMP and PRL also benefited from an invitation to Z\"{u}rich by
Prof. R. R. Ernst at an early stage of this work and stimulating comments of
Prof. A. Pines about our preliminary results. This work was performed at
LANAIS de RMN (UNC-CONICET), where the ferrocene single crystal was grown by
Rodrigo A. Iglesias. Financial support was received from Fundaci\'{o}n
Antorchas, CONICET, FoNCyT, CONICOR and SeCyT-UNC. The authors are
affiliated to CONICET.

\bigskip

.{\bf Figure 1}: {\bf Left panel}: Idealized representation of a
Polarization Echo experiment. The local polarization decay (dotted line) is
described by the Hamiltonian ${\cal H}+\Sigma $, where $\Sigma $ is small
enough for ${\cal H}$ to determine the spreading time $\tau _{mb}$. At time $%
t_{R}$ the dynamics is reversed. Further evolution of the polarization is
shown by the solid line. It develops a PE of amplitude $M_{PE}$. {\bf Right
panel: }Possible dependences of $M_{PE}$ as a function of $t_{R}$ . The
dotted line represents a system controlled by a strong relaxation rate ($%
\tau _{\phi }\ll \tau _{mb}$) while the solid line shows the opposite case.
The time scale $\tau $ in the plot corresponds to $\tau _{\phi }$ and $\tau
_{mb}$ respectively.

{\bf Figure 2}: Pulse sequence to control the dipolar spin dynamics after
creating a local polarization (see text). The solid and dotted lines show
the main and secondary amplitude polarization pathways.

{\bf Figure 3}: Attenuation of the polarization echo in ferrocene as a
function of $t_R$. The data were recorded using the pulse sequence
schematized in Fig. 2 with: $t_C=2ms$, $t_S=1ms$, $t_d=53\mu s$, and $%
n=1,2,8,16$. The attenuation is slowed down as the complexity reached by the
system is reduced. The lines represent Gaussian fittings with the
characteristic time $\tau _{mb}^n$ as the only free parameter. {\bf Inset}:
Characteristic time dependence on $n$.

{\bf Figure 4}: Same as in Fig. 3 but for cobaltocene sample. We used $%
t_C=85\mu s$, $t_S=150\mu s$, $t_d=85\mu s$, and $n=1,2,5,8,12$. There is a
clear crossover between a Gaussian and an exponential decay. This crossover
results from a change of the relative importance between $\tau _{mb}^n$ and $%
\tau _\phi $. The solid lines represent fittings of the whole set of data
with only three free parameters (see text).

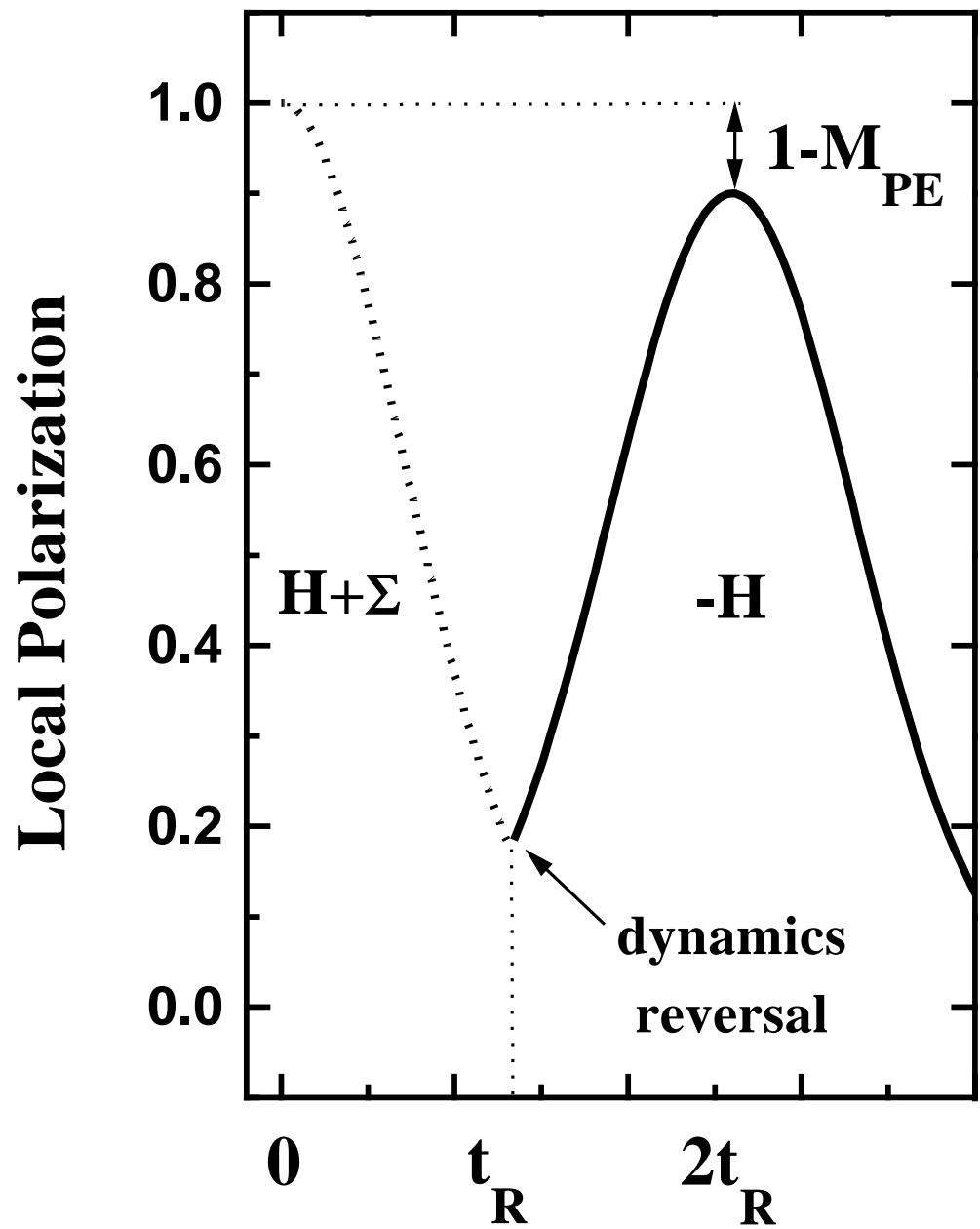
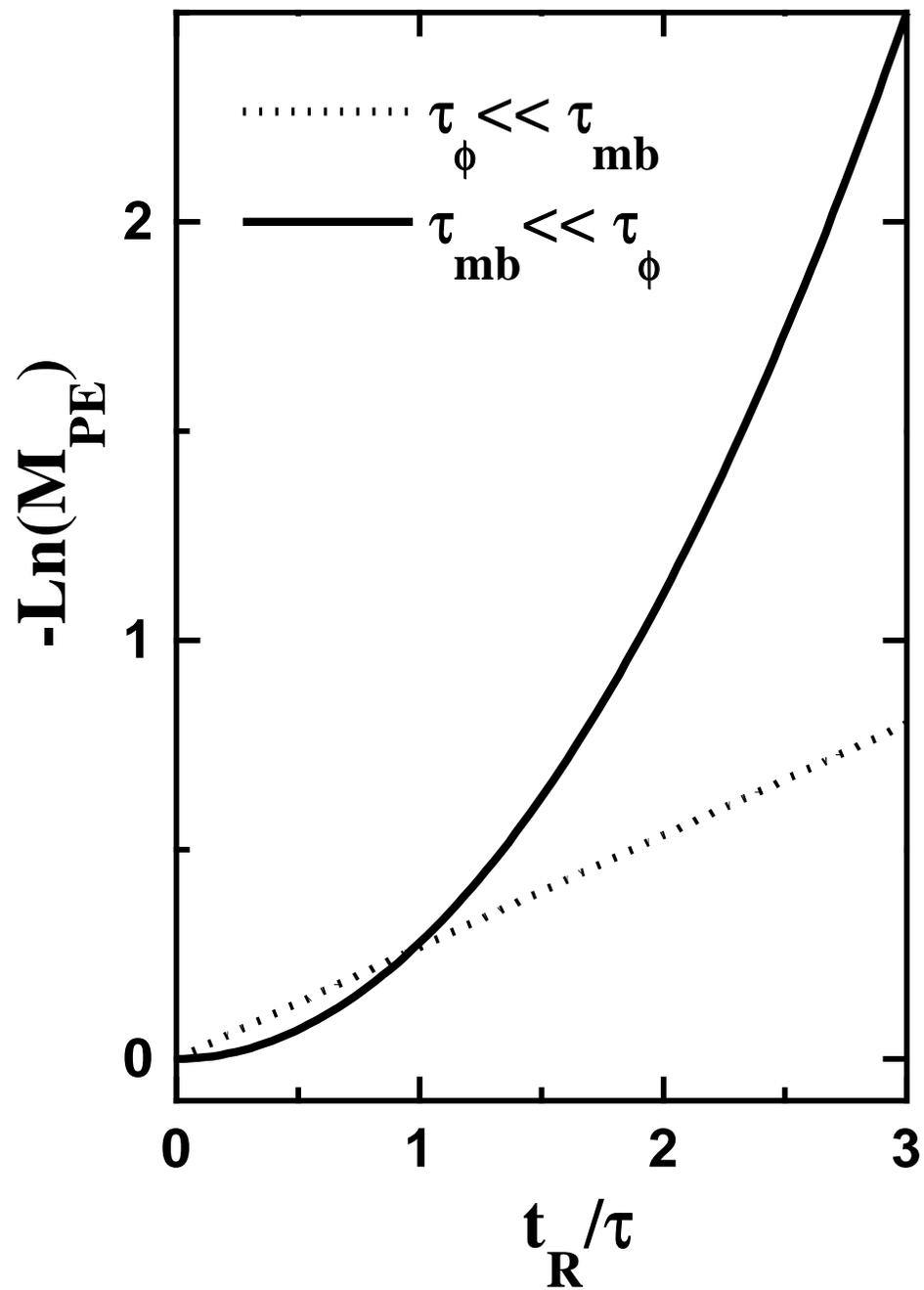
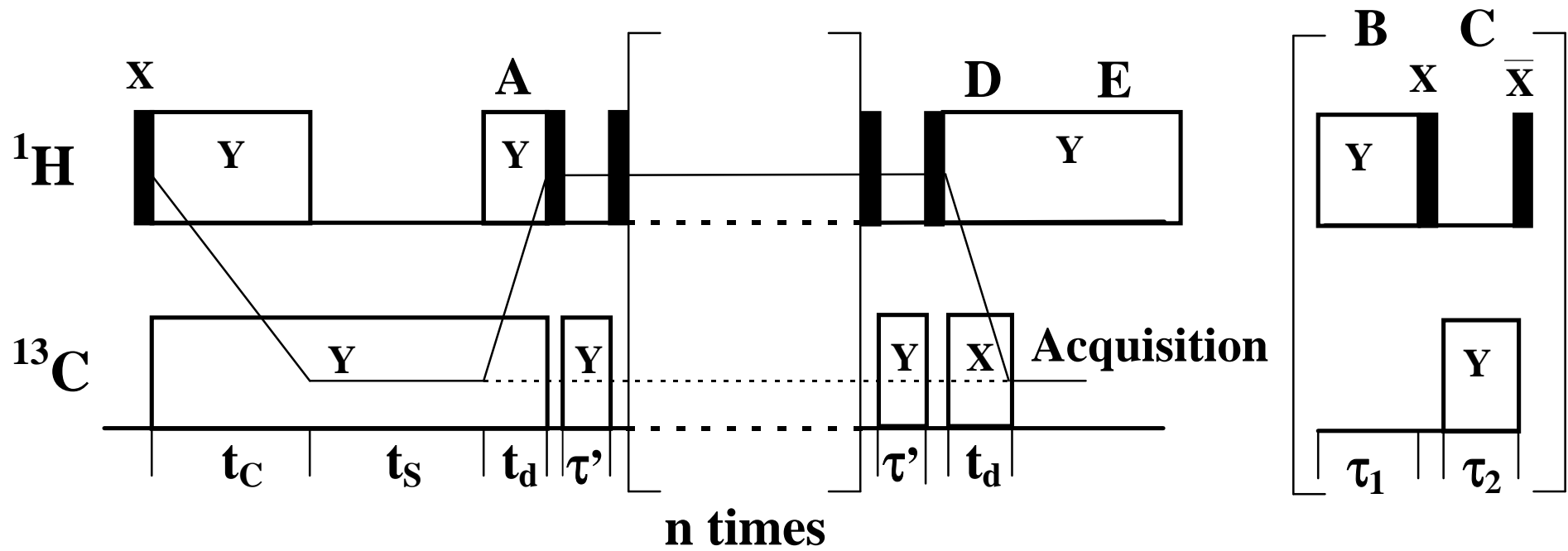
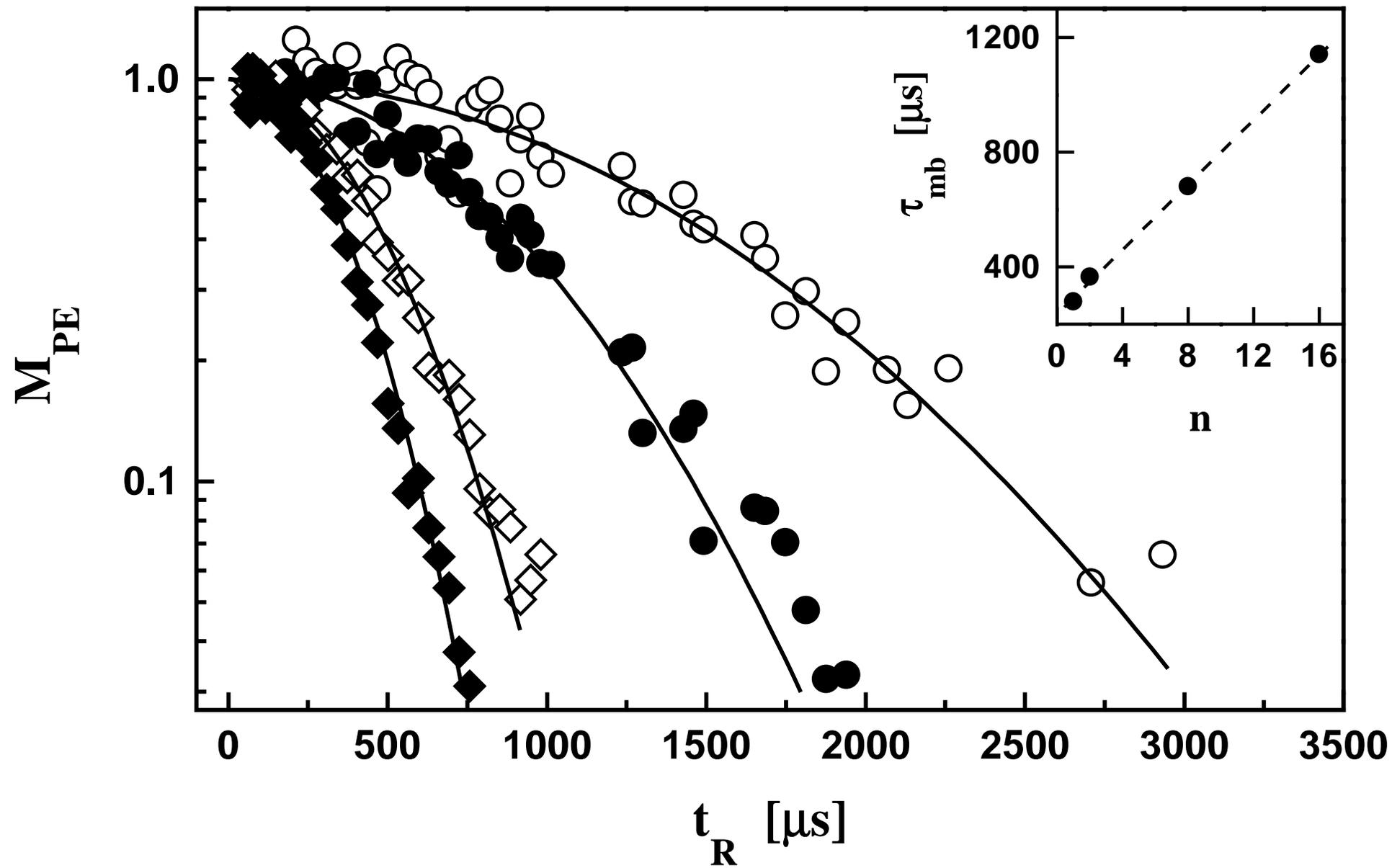
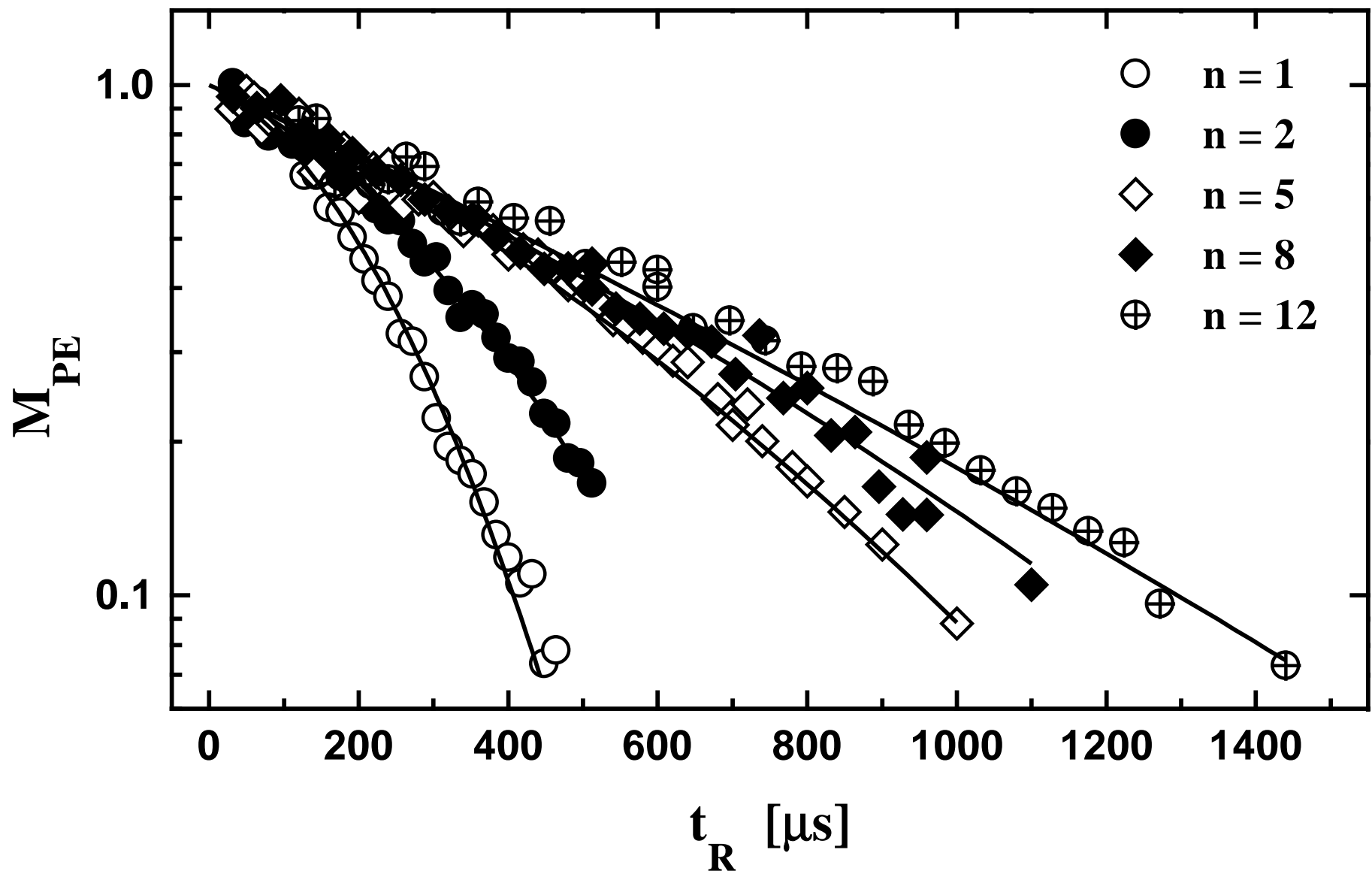
\end{document}